\documentclass[review]{elsarticle}

\usepackage{lineno,hyperref}
\usepackage{xcolor}
\modulolinenumbers[5]

\hypersetup{
	colorlinks=false,
	linkcolor = {blue!20!black},
	anchorcolor = {black},
	citecolor = {green!80!black},
	filecolor = {cyan!80!black},
	menucolor = {red!80!black},
	runcolor = {cyan!80!black},
	urlcolor = {magenta!80!black},
}

\usepackage{array}
\usepackage{multirow}
\usepackage{booktabs}
\usepackage{mathtools}
\usepackage{amsmath}
\usepackage{amssymb}
\usepackage[numbers]{natbib}
\biboptions{sort&compress}

\begin{document}
	
	\title{Data-mining of In-Situ TEM Experiments: on the Dynamics of Dislocations in CoCrFeMnNi Alloys}
	
	\author[1]{Chen Zhang}
	\author[1]{Hengxu Song}
	\author[2]{Daniela Oliveros}
	\author[3]{Anna Fraczkiewicz}
	\author[2]{Marc Legros}
	\author[1,4]{Stefan Sandfeld\corref{cor1}}\ead{s.sandfeld@fz-juelich.de}

	\cortext[cor1]{Corresponding author.}
	
	\address[1]{Institute for Advanced Simulations: Materials Data Science and Informatics (IAS-9), Forschungszentrum J\"{u}lich GmbH, 52425, J\"{u}lich, Germany}
	\address[2]{CEMES-CNRS, 29 Rue J. Marvig, 31055, Toulouse, France}
	\address[3]{\'Ecole des Mines de Saint \'Etienne, 42100 , Saint-\'Etienne, France}
	\address[4]{Faculty of Georesources and Materials Engineering, RWTH Aachen University, 52068, Aachen, Germany}

\begin{abstract}
High entropy alloys are a class of materials with many significant improvements in terms of mechanical properties as compared to ``classical'' alloys. The corresponding structure-property relations are not yet entirely clear, but it is commonly believed that the good mechanical performance is strongly related to dislocation interactions with the complex energy landscape formed due to alloying. Although in-situ Transmission Electron Microscopy (TEM) allows high-resolution studies of the structure and dynamics of moving dislocations and makes the local obstacle/energy ``landscape'' directly visible in the geometry of dislocations; such observation, however, are merely qualitative, and  detailed three-dimensional analyses of the interaction between dislocations and the energy landscape is still missing. In this work, we utilized dislocations as ``probes'' for the local energy maxima which play the role of pinning points for the dislocation movement. To this end, we developed a unique data-mining approach that can perform coarse-grained spatio-temporal analysis, making ensemble averaging of a considerable number of snapshots possible. We investigate the effect of pinning points on the dislocation gliding behavior of CoCrFeMnNi alloy during in-situ TEM straining and find that (i) the pinning point strength changes when dislocations glide through and (ii) the pinning point moves along the direction close to the Burgers vector direction. Our data-mining method can be applied to dislocation motion in general, making it a useful tool for dislocation research.	
\end{abstract}


	\maketitle
	
	\section{Introduction}\label{sec1}
	
	Multi-component complex alloys (MCCA) comprising near-equiatomic four or more elements have gained increasing attention due to their promising mechanical properties when compared to traditional single-phase alloys~\cite{yeh2004nanostructured, cantor2004microstructural, gorsse2018high, wu2014temperature, senkov2018development}. Hypotheses brought forward to explain these enhanced properties include severe lattice distortion, high entropy effect (from which the initial acronym HEA-High Entropy Alloy derives), and sluggish diffusion. These factors, specific to this new class of alloys may reinforce the regular strengthening mechanisms found in classical alloys and help achieve superior mechanical properties. The single-phase face-centered-cubic (fcc) CoCrFeMnNi alloy (also known as {``Cantor''} alloy~\cite{cantor2004microstructural}) is one of the most studied~\cite{zhang2021elevated,moretti2004depinning} and it exhibits high strength and toughness, even at cryogenic temperature~\cite{laplanche2016microstructure, otto2013influences}.
	
	The lattice friction, or Peierls barrier, is considered to be one of the key aspects involved in strengthening mechanisms. The magnitude of the lattice friction depends on the crystal lattice and atomic bonds and ultimately on the dislocation core structure. However, the dislocation core is usually planar in fcc material, consisting of two partial Shockley dislocations separated by a stacking fault. Therefore, lattice friction is often neglected when considering strengthening mechanisms in fcc metals and alloys. However, due to large lattice distortion and ubiquitous interactions among multiple principal atomic species, dislocations face a rugged atomic and energy landscape, possibly resulting in a new type of friction~\cite{ma2020unusual}. Through investigating thermal activation processes from stress-strain measurements with varying temperatures and strain rates for a family of equiatomic fcc single phase solid-solution alloys, Wu et al.~\cite{wu2016thermal} concluded that the Labusch-type solution strengthening mechanism, rather than lattice friction, governs the deformation behavior in equiatomic alloys. The solute strengthening in HEAs is likely to be associated with either short-range clustering or short-range ordering of solute atoms~\cite{zhang2020short, shen2021kinetic, he2021understanding, wu2021short}. However, Lee et al.~\cite{lee2020dislocation} did not observe any chemical heterogeneity or ordered structures in the CoCrFeMnNi alloy of sizes down to the resolution limits of $1\,\mathrm{nm}$ at cryogenic temperatures using experimental techniques based on electron diffraction, STEM-EDS (Scanning Transmission Electron Microscopy-Energy Dispersing Spectroscopy) and APT (Atom Probe Tomography). 
	
	Some researchers believe that strengthening is caused by phase transformation, since first-principle models anticipated a stable hcp phase at low temperatures~\cite{chen2020real, wang2020atomic}. Yu et al.\cite{chen2020real} observed that the fraction of the stronger hcp phase progressively increases during plastic deformation in the fcc phase in a Cantor-like $\mathrm{Cr}_{20} \mathrm{Mn}_{6} \mathrm{Fe}_{34} \mathrm{Co}_{34} \mathrm{Ni}_{6}$ alloy serving as the major source of strain hardening. However, the hcp phase has not been found in the equiatomic cantor alloy for the time being~\cite{lee2020dislocation}. 
	
	More people have recently recognized that local chemical fluctuation (LCF) is a common feature in HEAs and is considered to influence dislocation motion~\cite{varvenne2016theory, li2019strengthening, bu2021local, komarasamy2016anomalies}. Curtin et al.~\cite{varvenne2016theory} proposed a theory to explain the plastic yield strength for fcc HEAs that has been validated by molecular simulations on model Fe-Ni-Cr alloys. They consider each elemental component as a solute embedded in the effective matrix of the surrounding alloy, and deduced that the strengthening is mainly achieved due to dislocation interactions with the random local concentration fluctuations around the average composition. Ma et al.~\cite{li2019strengthening} constructed an atomic interaction potential for CrCoNi medium-entropy alloy and demonstrated that the local chemical ordering changed during processing and increased the ruggedness  of the local energy landscape and raised the activation barriers that govern dislocation activities. 
	
	These studies, mainly theoretical, therefore consider that strong pinning points are impeding the dislocation movement on their gliding planes in HEA, that these pinning points result from chemical and physical distortions of the lattice and are potentially dense enough to generate a general friction force on all dislocation movements. However, there are currently very few quantitative reports on pinning effects, because the interplay between energy and the very local chemical ordering is difficult to characterize and measure. Besides, small-scale simulations are spatially and temporally constrained, which does not allow them to grab a realistic and statistical view of such obstacles and their potential strength. As moving dislocations carry a shear that has an interatomic length scale, the landscape seen by one dislocation may differ substantially from the one seen by the following dislocations. This is especially true in alloys where local order may be built up or inversely deconstructed by successive shears of dislocations gliding on the same plane. As so, a moving dislocation, if one can follow its individual or collective motion, is the ultimate probe to explore such distorted landscapes as well as the vector of the on-going plastic deformation. We aim to characterize and quantify the essence of the pinning points encountered by moving dislocations at a larger time span and nanometer scale, based on experimental data from in-situ TEM straining.
	
	For that purpose, we observed representative pile-up of dislocations gliding on compact planes of a Cantor (CoCrFeMnNi equiatomic) alloy, then reconstructed the three-dimensional dislocation microstructures, which allowed us to detect the dislocation movement directly on a given glide plane. Next, based on the bent angle of dislocation when passing by the pinning point, we built a simple model to quantify the force of the pinning point and its evolution. In the last section, we carried out the spatio-temporal analysis of the local pinning points through coarse-graining of dislocation microstructure, in a way like the conventional scanning technique~\cite{friedbacher1999classification}. We show that dislocation lines face an evolving landscape as the deformation proceeds and that these local obstacles are probably at the root of hardening mechanisms in the Cantor alloy. The presented method may be generalized to other MCCA to assess the average ``friction stress'' that would result from these multiple local interactions.

	\section{Methods}\label{sec2}
	Below, we explain all the methods utilized in the analysis in detail. Section~\ref{subsection2-1} describes how the experiments were performed. Section~\ref{subsection2-2} is concerned with the three-dimensional reconstruction of dislocation lines from the projected, two-dimensional TEM images. Section~\ref{subsection2-3} introduces how the pinning point strength can be estimated, and Section~\ref{subsection2-4} derives and explains the spatio-temporal coarse graining of dislocation microstructures.
	\subsection{In-situ TEM straining experiment}\label{subsection2-1}	
	In-situ TEM straining experiments were carried out on a Cantor alloy processed at Mines Saint-Etienne, France, and prepared as electron-transparent stretchable samples following the method described in the previous work~\cite{oliveros2021orientation}. Additional videos can be found \href{https://www.youtube.com/channel/UCvaqUkpxoAnkb2Gj2Cz_OOw}{here}.
	Videos are generally acquired after strain increments are imposed to the sample on a Gatan straining holder model 671, operated at 96K. They are captured on a Megaview III SIS CCD camera, directly connected to a hard drive where they are stocked in mpeg4 format. 
	
	\subsection{3D reconstruction of dislocation microstructure from TEM images}\label{subsection2-2}
\begin{figure}[ht!]%
		\centering
		\includegraphics[width=0.75\textwidth]{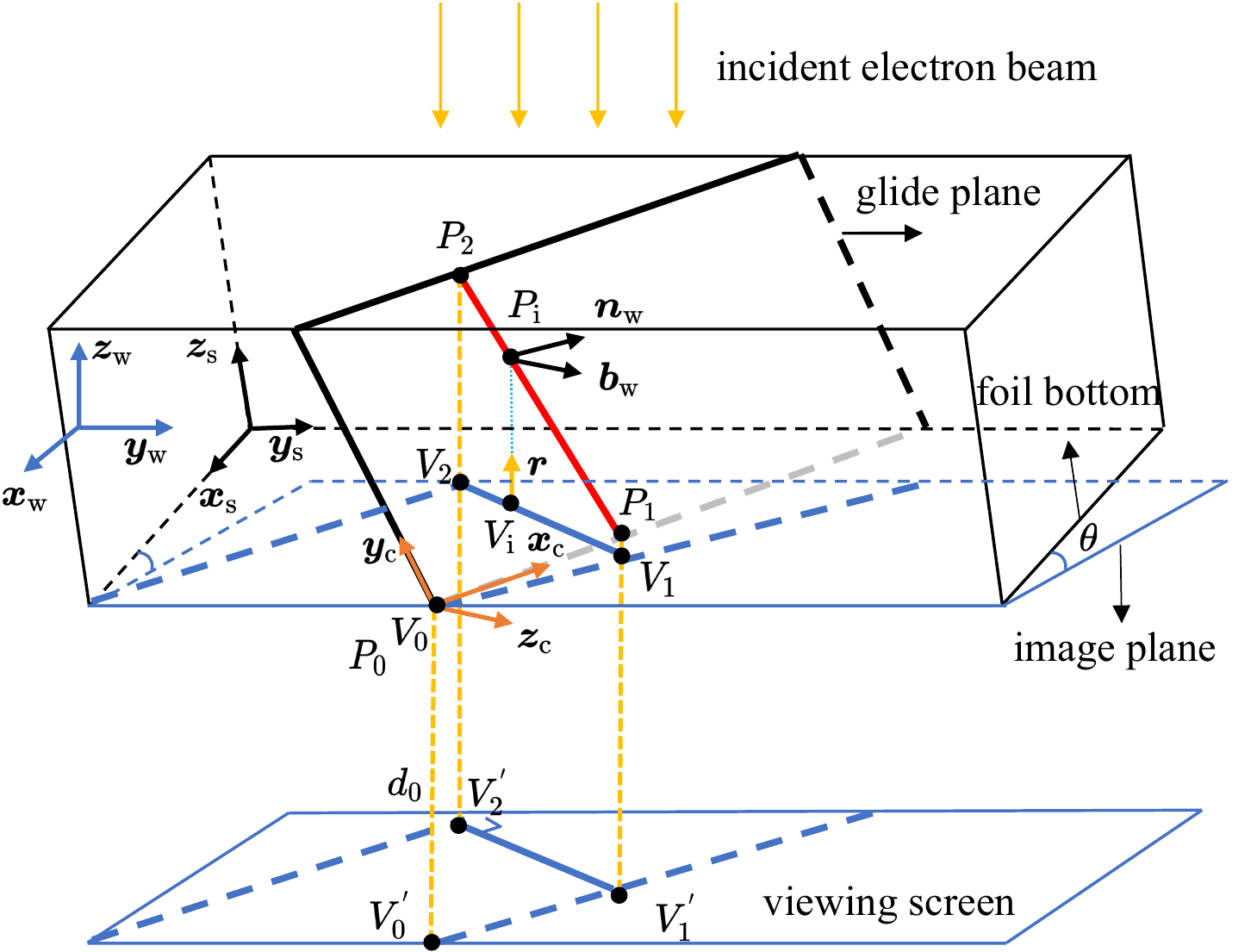}
		\caption{A dislocation ($P_{1}P_{2}$) in a foil is projected onto the image plane, becomes $V_{1}V_{2}$ and $V^{\prime}_{1}V^{\prime}_{2}$ in the viewing screen. Three different coordinate systems (crystal, sample, world) are involved, and they are correlated through Euler angles and tilt angle ($\theta$).}\label{fig1}
	\end{figure}

Dislocations as observed in the TEM images are essentially the projection of the real dislocation lines in $3 \mathrm{D}$ space onto the experimental screen plane. There are three orthogonal coordinate systems involved: 

\begin{enumerate}
	\item 
	The \textbf{World Coordinate System (WCS)}: is defined as the coordinate system of the TEM experiment bench. Usually, the sample is tilted by an angle $\theta$ along a certain axis (in our experiments, the tilt is along the $y$-axis) in WCS, as shown in Fig.\,\ref{fig1}.  

	\item 
	The \textbf{Sample Coordinate System (SCS)},
	is typically aligned with some of the axes of the sample geometry. For example, in Fig.\,\ref{fig1}, 
	the sample coordinate system $x_\mathrm{s}, y_\mathrm{s}, z_\mathrm{s}$ is defined along the sample edges.

	\item 
	The \textbf{Crystal Coordinate  System (CCS)} is defined along with the lattice basis. 
\end{enumerate}

	In the experiment, the CCS and SCS are linked through a transformation given by Euler angles that describe a sequence of rotations. In the material science community, there exists a certain ambiguity and inconsistency concerning the representation of and conversions between $3 \mathrm{D}$ rotations. While proper tensorial representations would be the most appropriate and unique descriptions, unfortunately, most of the time matrix notations are preferred that  need to be handled with care~\cite{rowenhorst2015}. A rotation matrix corresponding to some given Euler angles is unique only when the information about the following aspects is consistent and explicitly provided: 
(i) the ``handedness'' of the coordinate system (right-handed or left-handed), (ii) the ``type'' of the rotation (active or passive rotations), 
(iii) the Euler angle convention (such as the Tait-Bryan convention, Bunge convention) and (iv) the rotation sequence convention (intrinsic or extrinsic). As there is a certain amount of redundancy between (i) -- (iv) it is important that the information provided is consistent. A detailed discussion and python implementation details are given in \cite{ias9-cms}. Our TEM data were processed with the open-source software \textit{pycotem}~\cite{mompiou2021pycotem} where right-handed coordinate systems and active rotations are used. Furthermore, the software follows the Bunge convention with intrinsic rotations. Then, for a given Euler angle ($\phi_1$,\:$\Phi$,\:$\phi_2$), the corresponding rotation matrix reads:
\begin{equation}
\Re\left(z^{\prime \prime}\mid\phi_{2},\:x^{\prime}\mid \Phi,\:z \mid \phi_{1}\right)=R_{z}^{a}\left(\phi_{1}\right) \cdot R_{x}^{a}(\Phi) \cdot R_{z}^{a}\left(\phi_{2}\right)
\end{equation}

In the above equation, $\Re$ is an operator that executes the three comma-separated arguments from the right to the left. On the right-hand side of the equation is the chain of rotation matrices with $ R_z^{a}$ and $R_x^{a}$ being the active rotation matrices along the $z$ and $x$-axis, respectively. The dot indicates the regular matrix-matrix multiplication that again results in a matrix.

Having clarified the Euler angle conventions, we are able to calculate dislocation-related information such as the Burgers vector and slip plane normal in the SCS, and finally in the WCS after further tilting. The detailed values are shown in Tab.\,\ref{tab1}.  Having all the information in WCS, reconstructing the dislocation microstructure is essentially identical to projecting dislocation lines from the viewing screen back to the slip plane, as shown in  Fig.\,\ref{fig1}. For a given point $\boldsymbol{P}_\mathrm{i}$ on the slip plane, its projection on the viewing screen is $\boldsymbol{V}_\mathrm{i}$ where 
\begin{equation}
\boldsymbol{P}_\mathrm{i}=\boldsymbol{V}_\mathrm{i}+\boldsymbol{r} d_\mathrm{i}
\end{equation}
where $\boldsymbol{r}$ is a unit vector along the reverse electron beam direction/global $z$ direction, i.e., [0 0 1]. On the slip plane, we have
\begin{equation}
\left(\boldsymbol{P}_\mathrm{i}-\boldsymbol{P}_\mathrm{0}\right) \cdot \boldsymbol{n}_\mathrm{w}=0
\end{equation}
where $\boldsymbol{P}_\mathrm{0}$ is any point on the slip plane, and $\boldsymbol{n}_\mathrm{w}$ is the slip plane normal under WCS. Combining the above two equations, we get 
\begin{equation}
d_\mathrm{i}=\frac{\boldsymbol{n}_\mathrm{w} \cdot\left(\boldsymbol{P}_\mathrm{0}-\boldsymbol{V}_\mathrm{i}\right)}{\boldsymbol{n}_\mathrm{w} \cdot \boldsymbol{r}}
\end{equation}

For the convenience of the calculation, we let $\boldsymbol{P}_\mathrm{0}$ overlap with $\boldsymbol{V}_\mathrm{0}$ as the origin on the viewing screen. After solving $d_\mathrm{i}$, any point $\boldsymbol{V}_\mathrm{i}$, which are the dislocation polygon's vertices, can be projected back to the slip plane.

For all dislocation coordinates $\boldsymbol{P}_\mathrm{i}$ in 3D, we further project them onto the slip plane where $\boldsymbol{P}_\mathrm{0}$ is taken as the origin, the Burgers vector $\boldsymbol{b}_\mathrm{w}$ is taken as the local $x$ axis and slip plane normal $\boldsymbol{n}_\mathrm{w}$ as local $z$-axis, thus we get dislocation coordinates on the slip plane $\boldsymbol{P}^{\rm{sp}}_\mathrm{i}$:
\begin{equation}
\boldsymbol{P}^{\rm{sp}}_\mathrm{i}=
\left[\boldsymbol{b}_\mathrm{w} \cdot \boldsymbol{P}_\mathrm{i} \boldsymbol{P}_\mathrm{0},  \boldsymbol{n}_\mathrm{w}  \times \boldsymbol{b}_\mathrm{w} \cdot \boldsymbol{P}_\mathrm{i} \boldsymbol{P}_\mathrm{0}, 0\right]
\end{equation}
	\begin{table}[ht!]
		\begin{center}
			\caption{Definition and value of symbols}\label{tab1}%
			\small
			\begin{tabular}{@{}lll@{}}
				\toprule
				Symbol & Definition  & value\\
				\midrule
				$\left(\phi_{1}, \Phi, \phi_{2}\right)$ & Bunge Euler angles & $\left(69.3^{\circ}, 39.5^{\circ},-68.1^{\circ}\right)$ \\
				$\boldsymbol{r}$ & Reversed electron beam direction, in WCS & $(0,0,1)$ \\
				$\boldsymbol{n}_\mathrm{c}$ & Slip plane normal, vector, in CCS & $[\bar{1},\bar{1}, 1]$ \\
				$\boldsymbol{n}_\mathrm{w}$ & Slip plane normal, unit vector, in WCS & $[-0.125, -0.745, 0.655 ]$  \\
				$\boldsymbol{b}_\mathrm{c}$ & Burgers vector, vector, in CCS & $(-1, 1, 0)$ \\
				$\boldsymbol{b}_\mathrm{w}$ & Burgers vector, unit vector, in WCS & $(-0.499, 0.618, 0.607)$ \\
				$\theta$ & Tilt angle, along $y_\mathrm{w}$ axis & $2.5^{\circ}$ \\

				\bottomrule
			\end{tabular}
		\end{center}
	\end{table}
	
	\subsection{Estimation of pinning point strength}\label{subsection2-3}
	In this section, we utilized the simple line tension model to estimate the strength of the pinning point. The line tension of a dislocation segment with the mixed character reads: 
		\begin{equation}
		T=E_{\rm{el}}(\alpha)+\frac{\mathrm{d}^{2} E_{\rm{el}}(\alpha)}{\mathrm{d} \alpha^{2}}
	\end{equation}
	where $\alpha$ is angle between the Burgers vector and the line direction, and ${E}_{\rm{el}}(\alpha)$ is the elastic energy of the dislocation according to~{\cite{hull2011introduction}}:
\begin{eqnarray}  
		E_{\rm{e l}}(\alpha)&=&\left[\frac{G b^{2} \sin ^{2} \alpha}{4 \pi(1-v)}+\frac{G b^{2} \cos ^{2} \alpha}{4 \pi}\right] \ln \left(\frac{R}{r_{0}}\right)\\\nonumber
		&=&\frac{G b^{2}\left(1-v \cos ^{2} \alpha\right)}{4 \pi(1-v)} \ln \left(\frac{R}{r_{0}}\right)
	\end{eqnarray}	
Here, $\mathrm{R}$ is the outer radius (distance from the dislocation core) and $\mathrm{r_{0}}$ is the core radius, which take the values $\mathrm{R}/\mathrm{r}_{0}=4$ and $G=84.85\,\mathrm{GPa}$ in our calculations.

	\begin{figure}[ht!]%
		\centering
		\includegraphics[width=0.7\textwidth]{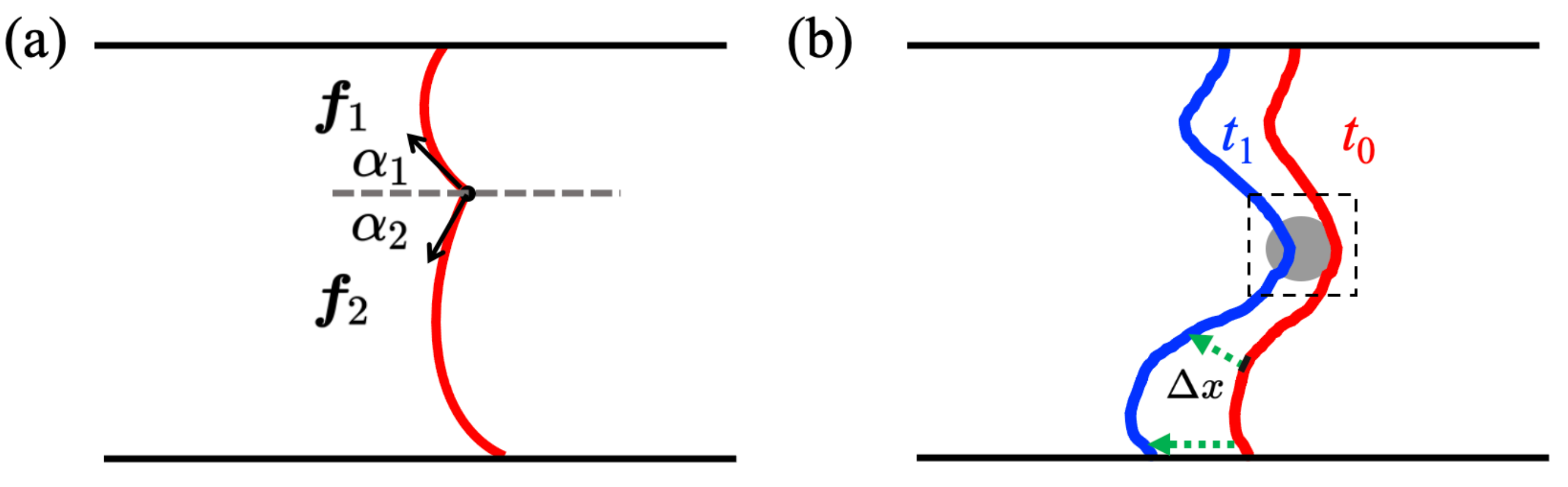}
		\caption{Schematic of dislocation geometry around the local pinning point. (a) Simplified model of dislocation pinned at the obstacle.  (b)Two different configurations of dislocation at different times. Within the time span $\Delta t$, the dislocation segment travels a distance $\Delta x$.}\label{fig2}
	\end{figure}
	
As shown in {Fig.\,\ref{fig2}a}, the dislocation is pinned by an obstacle. Close to the pinning point where the force vectors originated from line tensions are $\boldsymbol{f_{1}}$ and $\boldsymbol{f_{2}}$, respectively, the dislocation line tangents form the angles $\alpha_{1}$ and $\alpha_{2}$ with the horizontal direction (indicated by the dashed line). We assume that the pinning point only exhibits an interaction strength along the horizontal direction. Then, based on the force balance, we have
\begin{equation}
		 \left\lVert \boldsymbol{f}_{\rm{obs}} \right\rVert= \left\lVert \boldsymbol{f}_{1}\right\rVert \cos \alpha_{1}+\left\lVert \boldsymbol{f}_{2}\right\rVert \cos \alpha_{2}
	\end{equation}

\subsection{Spatio-temporal coarse graining of dislocation microstructures}
\label{subsection2-4}
Dislocations nucleate and evolve under the influence of external stress, due to interaction with other dislocations, as well as due to interactions with inclusions, phase or grain boundaries. 
While the exact, mathematical description of all individual dislocations as oriented curves  might seem like the most accurate representation, it is nonetheless difficult to understand their collective behavior from there.
Coarse-grained representations of dislocation microstructures provide averaged information on a larger length scale that is defined by the averaging voxel size. Thereby, the goal is that only the ``relevant'' information are preserved while noise and irrelevant details are averaged out \cite{Song2021_MSMSE29}. The ``relevant'' information is represented in form of the density and density-like field variables of the Continuum Dislocation Dynamics (CDD) theory~\cite{hochrainer2014continuum}. These field variables represent features of dislocation microstructures, e.g., the geometrically necessary dislocation density is related to a plastic strain gradient~\cite{sandfeld2010numerical}, and the curvature represents how strongly dislocations are bent at/around inclusions~\cite{sandfeld2015microstructural}. 
The detailed mathematical derivation and numerical implementation of the coarse-grained dislocation quantities can be found in~\cite{sandfeld2015microstructural}. Here we briefly summarize the formulations of the two quantities we utilized in our analysis: the coarse grained dislocation density $\rho$ and the coarse grained dislocation curvature density $q$. For this, we start with the discrete variants of these two quantities that are defined for the mathematical line $\boldsymbol{c}$:

\begin{equation}
\rho_{\boldsymbol{c}}(\boldsymbol{r}) = \int^{L_{c}}_{0}\delta(\boldsymbol{c}(s)-\boldsymbol{r}) \mathrm{d}s
\end{equation}

\begin{equation}
q_{\boldsymbol{c}}(\boldsymbol{r}) = \int^{L_{c}}_{0}\delta(\boldsymbol{c}(s)-\boldsymbol{r}) \frac{\mathrm{d}^2\boldsymbol{c}}{\mathrm{d}s^2} \cdot \left( \frac{\mathrm{d}\boldsymbol{c}}{\mathrm{d}s}\times \boldsymbol{n} \right)\mathrm{d}s, 
\end{equation}
There, $L_{c}$ is the total length of the dislocation curve $\boldsymbol{c}(s)$, $\delta()$ is the Dirac delta function, and $\boldsymbol{n}$ is the slip plane normal. We average these quantities in a domain $\Omega$ of size $V$ centered at $\boldsymbol{r}$ through an averaging operator defined as 
\begin{equation}
\left\langle \circledast \right\rangle_{\Omega,\boldsymbol{r}} = (1/V) \int_{\Omega} \circledast\, \mathrm{d}^{3}r,
\end{equation}
where the symbol $\circledast$ is a placeholder for the quantity to be averaged. Using this, we get the average fields:
\begin{equation}
\rho(\boldsymbol{r}) = \frac{1}{V}\sum_{c}\int_{\mathcal{L}_{c}^{\Omega,\boldsymbol{r}}} 
\!1\;  \mathrm{d}\ell
\end{equation}
\begin{equation}
q(\boldsymbol{r}) = \frac{1}{V}\sum_{c}\int_{\mathcal{L}_{c}^{\Omega,\boldsymbol{r}}}\frac{\mathrm{d}^2\boldsymbol{c}}{\mathrm{d}\ell^2}  \cdot \left( \frac{\mathrm{d}\boldsymbol{c}}{\mathrm{d}\ell}\times \boldsymbol{n} \right)\mathrm{d}\ell
\end{equation}
where $\mathcal{L}_{c}^{\Omega,\boldsymbol{r}} \subset \boldsymbol{c}$ denotes a section of line $\boldsymbol{c}$ contained inside the averaging volume $\Omega$ whose center is at $\boldsymbol{r}$. 
Numerically, we discretize the domain into many voxels/pixels whose center $\boldsymbol{r}_{i}$ has the coordinates $(x_{i}, y_{i}, z_{i})$. Each voxel defines an averaging sub-domain $\Omega_{i}$:
\begin{equation}
\Omega_{i} = \left[x_i - \frac{1}{2}\Delta x, x_i +  \frac{1}{2}\Delta x\right] \times ...\times  \left[z_i - \frac{1}{2}\Delta z, z_i +  \frac{1}{2}\Delta z\right] 
\end{equation}
with subvolume $V_{i} = \Delta x\Delta y\Delta z$. Therefore, for each voxel/pixel, we can calculate the coarse grained dislocation density $\rho_{i}$ and coarse grained curvature density $q_{i}$. The average curvature is then calculated as 
\begin{equation}
\kappa_{i} = q_{i}/\rho_{i} 
\end{equation}

The above coarse-graining spatially averages given dislocation microstructures. At the same time, dislocation microstructure evolves due to the change of the external loading, mutual interactions, or internal features, e.g., diffusion of precipitates. In this paper, the goal is to identify the obstacles, the concomitant energy landscape, and their evolution using the coarse-grained curvature and velocity. To analyze slow or stationary details, we carry out further temporal averaging, which altogether results in spatio-temporal coarse graining of dislocation microstructures.
For a given frame picture at time $t=j$, we can get the average dislocation curvature $\kappa^{j}_{i}$ and the average dislocation velocity $v^{j}_{i}$ (details can be found in section~\ref{subsection3-3}). The temporal average can be calculated through the weighted average by the coarse-grained dislocation density, which reads as follows:
\begin{equation}
\bar{\kappa}_{i} = \frac{\sum^{n}_{j=0} \rho^{j}_{i}\kappa^{j}_{i}} {\sum^{n}_{j=0}\rho^{j}_{i}}, \quad \bar{v}_{i} = \frac{\sum^{n}_{j=0} \rho^{j}_{i}v^{j}_{i}} {\sum^{n}_{j=0}\rho^{j}_{i}}
\end{equation}

	\section{Results and discussion}\label{sec3}
	In the following, we start our data-mining approach by describing the steps required for the three-dimensional reconstruction of the dislocation structure based on the recorded in-situ TEM videos from which images are extracted. We analyze the data by extracting the geometrical information about the local shape of dislocations. In the end, through coarse-grained quantities, we carry out a `scan' of the whole slip plane, utilizing dislocations as probes to extract previously inaccessible information about the local pinning in this representative HEA.
	
	\subsection{3D reconstruction of the dislocation structure}\label{subsection3-1}
	
	The first and most important input (besides the TEM images themselves) for reconstructing the $3 \mathrm{D}$  pile-up structure is the knowledge of the exact Euler angles and tilt convention used in the experiment. We summarize these steps as the ``3D reconstruction technique'', see section \ref{subsection2-2} for details). Based on this, it is possible to map the dislocations from the TEM image projection back to the distribution along their specific slip plane. 
	
\begin{figure}[h!]%
		\centering
		\includegraphics[width=\textwidth]{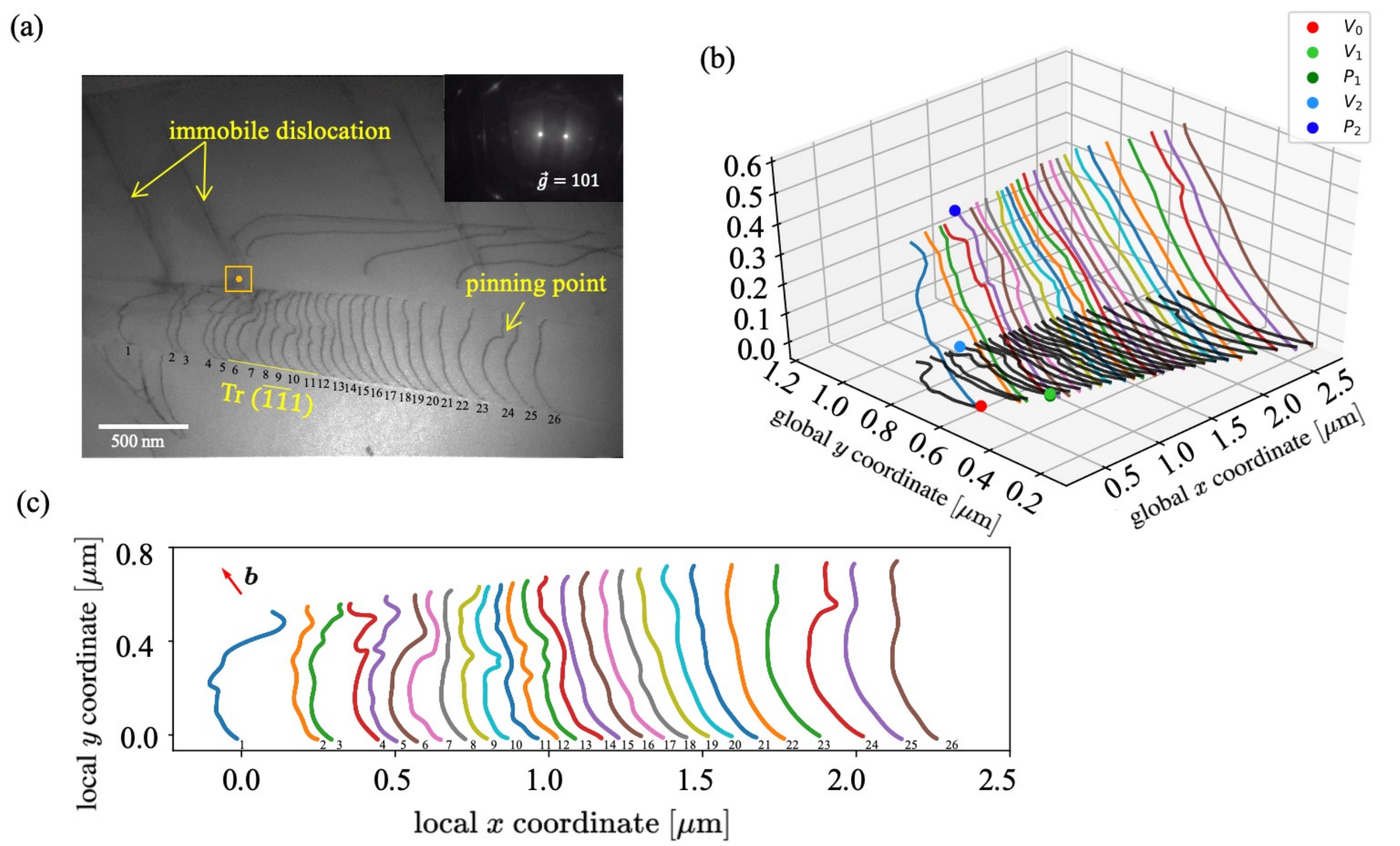}
           		\caption{Frame $0561(t=75\,\mathrm{s})$ of piled-up dislocations in CoCrFeMnNi HEA during in-situ TEM straining at $96\,\mathrm{K}.$ (a) A $3 \mathrm{D}$ plot of plied-up dislocations in world coordinate system. (b) Configuration of dislocation piled-up on image plane with the corresponding diffraction pattern on upper right. (c) Configuration of the dislocation piled-up {on} the $\left[\bar{1},\bar{1}, 1\right]$ glide plane. The red arrow represents the direction of the Burgers vector $\left(-1,1, 0\right)$.}\label{fig3}
	\end{figure}

	In this work, the initial video in which dislocations are frequently pinned has a duration of around 4 minutes. During the pre-processing, we utilized a clearly visible reference point marked as $\boxdot$ in Fig~\ref{fig3}(a) to stabilize the video ~\cite{ias9-cms}. The stabilized video has been accelerated to 58 seconds while keeping the key frames (30 frames per second), yielding 1743 images in total. Unless otherwise stated, all the following discussions correspond to the real-time 4-minute video time frame. The open-source software $labelme$~\cite{wada2019labelme} was used to mark (or ``label'') each dislocation in the pile-up, i.e., draw a thin line on top of the dislocation and thereby creating polygons from which the points'  coordinates can then be used. The polygon data of  dislocations are indexed from the left to the right, and for each dislocation, the coordinates of the polygon are indexed from the bottom to the top. For all 1740 frames in the 58 second video, we labelled at least one frame per second, resulting in 303 data records where each record contains the full information of the dislocation structure of a frame.
	
	{Through the 3D reconstruction method explained in section~\ref{subsection2-2}, we are able to reconstruct the 3D dislocation microstructures based on dislocations in a TEM image (Fig.~\ref{fig3}a). The reconstructed 3D microstructure is shown in Fig.~\ref{fig3}b and is further projected on the slip plane, as shown in Fig.\,\ref{fig3}c.} It can be seen that the dislocations have varying ``height'', i.e. the vertical distance between the dislocation end points in the $z$-direction. Since a single dislocation (which does not intersect with other crystallographic defects) can only end at the free surface of the sample, the varying ``height'' therefore means the thickness of the sample changes. This variation is typical of electrochemically polished samples. In the present example, the calculated sample thickness varies around 460$\sim$560nm, a rather large value that has been validated using FIB cross-sections of the observed TEM samples.
	
	 The shape of a dislocation is influenced by the interactions with other dislocations, the free-surface effect, the external loading, and local pinning effect. The first two contributions do not vary much when we study an isolated dislocation at the tail of the pile up. The resolved shear stress on a dislocation is estimated through the simple dislocation circle model $\tau=\mu b/R$ where $R$ is the radius of the fitted dislocation circle and $\mu$ is the shear modulus ($84.85\,\mathrm{GPa}$~\cite{otto2013influences}) and $b$ is the magnitude ($0.254\,\mathrm{nm}$) of the Burgers vector. For dislocations at the tail of the pile up, the estimated resolved shear stress is $27.57\pm 3.81\,\mathrm{MPa}$, where the error margin is the standard deviation of estimations over time. Considering that the variation is rather small, we therefore exclude the effect of the external stress on the evolution of dislocation shapes, i.e, change of the dislocation shape mainly results from the local pinning points.
	
	\subsection{Identification of the local pinning point and evolution of strength}\label{subsection3-2}
	The local pinning point, unlike inclusions of different phases, is not visually observable in TEM images. However, it has a clear effect on the movement of dislocations: it reduces the dislocation mobility locally close to the pinning point, which eventually results in the bowing-out of dislocations in the opposite direction to the gliding direction (as seen in Fig.\,\ref{fig3}a). Therefore, the dislocation can be used as a probe of the local pinning point: the high curvature regions along the dislocation are potential locations for the pinning point.
	
For the identification of the local pinning point, we first artificially identify three possible pinning points $P_{1}, P_{2}$, and $P_{3}$ in the TEM video, where dislocations are always clearly bent when traveling around these points. Next, for each pinning point, we calculated the dislocation curvature (along the dislocation line) when dislocation moves around the effective zone of the local pinning point as indicated by the dashed square in Fig.\,\ref{fig2}b. It should be noted that the maximum curvature of a dislocation is sometimes located at the end points. This is caused by the ``friction effect'' resulting from the surface oxides, combined with image forces from the free surface. The local pinning point region is the location where the dislocation has a local maximum curvature inside the effective zone. Overall, we have 19, 11, and 50 configurations when dislocation gets pinned in the effective zone of $P_{1}, P_{2}, P_{3}$, respectively.  We further categorized the dislocation index in these configurations, resulting in the collected data shown in Tab.\,\ref{tab2}. For example, for $P_{1}$, dislocation \#15 was pinned twice, \#16 was pinned nine times, and \#17 was pinned eight times when it passed.

\begin{table}[ht!]
	\caption{%
		Overview of the data collected for the possible pinning points $P_{1}, P_{2}$, and $P_{3}$. 
		Below shows the dislocation index and the number of configurations for each pinning point.
	}
	\begin{center}
		\def\arraystretch{1.1} 
		\small
		\begin{tabular}%
		{%
			!{\extracolsep{2em}}
			>{\centering}p{1.1cm}@{}>{\centering}p{1.4cm}  
			>{\centering}p{1.1cm}@{}>{\centering}p{1.4cm}  
			>{\centering}p{1.1cm}@{}>{\centering\arraybackslash}p{1.4cm}%
		}
			\hline 
			\multicolumn{2}{c}{\textbf{pinning point $\boldsymbol{P_1}$}} & 
			\multicolumn{2}{c}{\textbf{pinning point $\boldsymbol{P_2}$}} & 
			\multicolumn{2}{c}{\textbf{pinning point $\boldsymbol{P_3}$}}  \\
			\footnotesize dislocation ID & \footnotesize number of configs. & \footnotesize dislocation ID & \footnotesize number of configs. & \footnotesize dislocation ID & \footnotesize number of configs. \\
			\cline{1-2} \cline{3-4} \cline{5-6}
			\#15 & 2 & \#18 & 6 & \#22 & 2 \\
			\#16 & 9 & \#19 & 1 & \#23 & 39 \\
			\#17 & 8 & \#20 & 4 & \#25 & 4 \\
				&   &      &   & \#26 & 5 \\
			\hline\\
		\end{tabular}		
	\end{center} 
	\label{tab2}%
\end{table}
	
\begin{figure}[htp!]%
		\centering
		\includegraphics[width=0.9\textwidth]{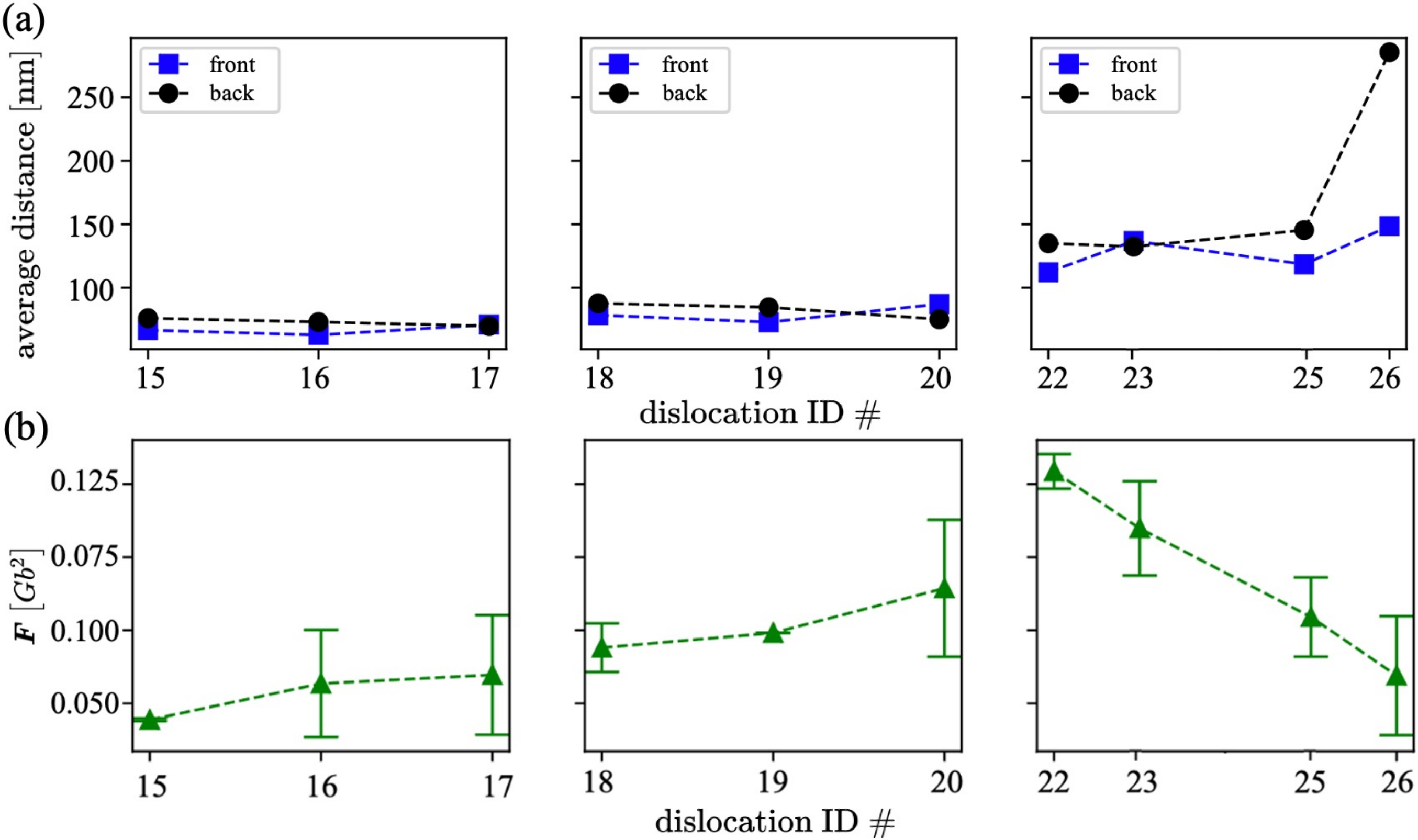}
		\caption{(a) Variation of the distance between the pinned dislocation and its adjacent dislocations (front  and back). (b) Strength of the local pinning point estimated by different dislocations passing-by. The rough locations are marked with cyan boxes in Fig.\,\ref{fig5}b. From left to right, the corresponding points are $P_{1}$, $P_{2}$, and $P_{3}$; the time spans corresponding to their strength changes are 1.3-14.2$\,\mathrm{s}$, 14.4-24.5$\,\mathrm{s}$, and 0-23.6$\,\mathrm{s}$, respectively.}\label{fig4}
	\end{figure}

The strength of the local pinning point can be estimated by the simple line tension model shown in Fig.\,\ref{fig2}a. In summary, the bending angle of the dislocation is directly related to the strength of the local pinning point; a smaller angle corresponds to a stronger local pinning point. In the previous section, we excluded the effect of the external stress on the dislocation shape evolution. However, since the analyzed dislocation comes from the dislocation pile up, the dislocation interactions can also alter the dislocation shape and influence the bent angle. To clarify the effect of dislocation interactions, we measure the average distance of a pinned dislocation with its two nearest neighboring dislocations (front and back). 

It can be seen in Fig.\,\ref{fig4}a that for $P_{1}$ and $P_{2}$, that when different dislocations get pinned, the distance between their neighboring dislocations does not significantly change. This indicates that the contribution from dislocation interactions does not change, i.e., the shape change (or the change of the bending angle) essentially results from the evolution of the pinning point strength. We therefore calculated the pinning point strength evolution based on the evolution of the bending angle;  results are shown in Fig.\,\ref{fig4}b. 
For $P_{1}$ and $P_{2}$, the pinning point strength slightly increases when dislocations pass by. For example, the pinning point strength calculated based on dislocation \#16 is larger than that calculated by dislocation \#15 which passes the pinning point prior to \#16 getting pinned at $P_{1}$. By contrast, the pinning point strength at $P_{3}$ decreases, especially for dislocation \#22, \#23 and \#25 whose distance to neighboring dislocations does not really change. 

This finding highlights the difference in the strengthening mechanism between HEA and conventional solute-hardened alloys: In solute-hardened alloys, the solid solution hardening comes from the interactions between dislocations and the solution stress field (together with the solution strength when a dislocation cuts through the solute). The effective strength of these solutes is generally constant or decreases after being sheared. However, in the studied HEA, the effective strength of the pinning points (lattice distortion from local ordering) evolves in a `random' manner. Some pinning points are hardened while others are weakened. The strength of the pinning point changes all the time in the process of dislocation gliding.
	
\subsection{Spatio-temporal analysis of the local pinning points through coarse graining of dislocation microstructure}\label{subsection3-3}

So far, we have identified the local pinning points by sight and studied the evolution of their strength through the change of the bending angle when dislocations pass by. In this section, we perform a spatio-temporal coarse graining of the dislocation microstructure, which will simultaneously reveal the local pinning points. The idea is as follows: Dislocations can be regarded as probes that are sampling the local material configuration. As a ``reaction'' to the local atomic arrangement, dislocations in an external stress field change their shape, which is shown in the local line curvature. Additionally, the local velocity changes. These two aspects are the ``signal'' received through the dislocation. We also notice that an ``action'' where dislocations interact with the obstacles also changes the atomic configuration. 

By scanning the whole slip plane using dislocations as probes and focusing on the above two ``signals'', we get a clear picture of the whole area and are able to detect the local pinning points. Resolution is always a key parameter in all types of scanning techniques. In our case, the resolution is defined as the size of the voxels/pixels that are used to discretize the whole slip plane. Within each pixel, the signals (curvature and velocity) are averaged both spatially and temporally. The spatial average of the signals of all considered dislocations gives rise to the so-called coarse-grained dislocation fields, which are effective measurements of complex dislocation microstructures~\cite{sandfeld2015microstructural}. 
The temporal average is the ensemble average of all scanned realizations. It is, however, not a simple arithmetic average of all realizations but rather an average where the contributions of individual averaging volumes are weighted by coarse-grained dislocation density values. That means, only those voxels that contain dislocations are considered. A detailed explanation of the spatio-temporal coarse-graining of the dislocation field quantities and their respective numerical calculation procedures are explained in the methods section \ref{subsection2-4}.

	\begin{figure}[ht!]%
		\centering
		\includegraphics[width=1.0\textwidth]{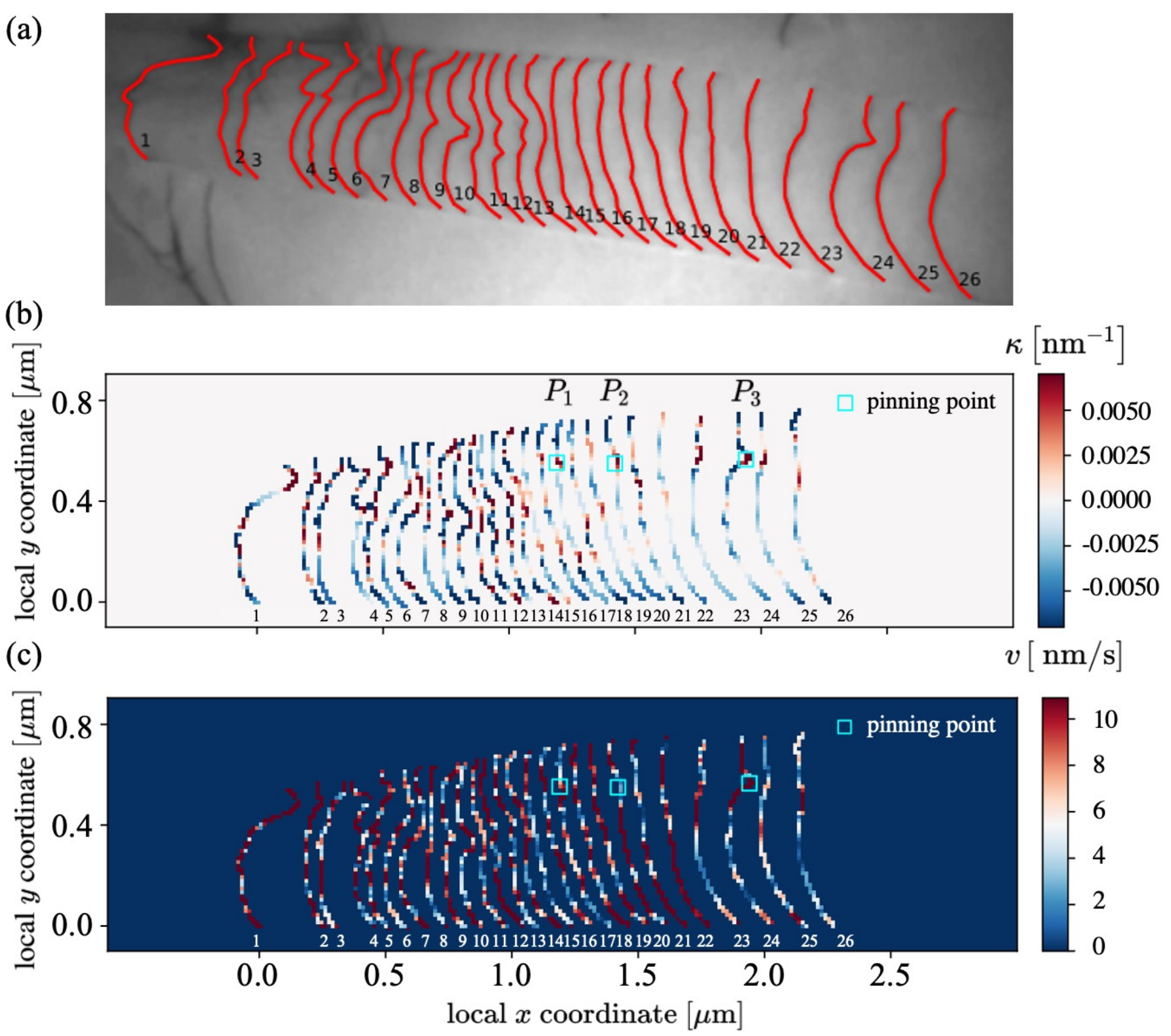}
		\caption{%
			Example of scanning using dislocation as the probe; (a) Labelled dislocations in the original TEM image, (b) Coarse-grained dislocation curvature with a pixel size of $15\,\mathrm{nm}$. (c) Coarse-grained dislocation velocity calculated based on data in two frames.}
		\label{fig5}
	\end{figure}
	
	An example of such a scanned slip plane can be found in Fig.\,\ref{fig5}, where a coarse-graining resolution of $15\,\mathrm{nm}$ was prescribed. Decreasing the resolutions any further does not yield significantly different results. The identified dislocations in the TEM images, shown in Fig.\,\ref{fig5}a, are first reconstructed into 3D, then mapped onto their slip plane. The coarse grained curvature for this snapshot is shown in Fig.\,\ref{fig5}b. There we see that the curvature varies along each single dislocation, which clearly reflects the rough and wavy feature of the dislocation line in HEA. The coarse-grained curvature has both positive and negative values depending on the bending direction of dislocations. In our calculation, if the dislocation is bent against the moving direction (i.e., towards the positive $x$ direction), the curvature is positive. We also marked the three pinning points that were manually identified in the previous section (cf. bottom panel of Fig.~\ref{fig5}). It can be clearly seen  that the coarse-grained curvature has high positive values around those three points. 
	
	{The calculation of the dislocation velocity is schematically shown in Fig.~\ref{fig2}b: based on the polygon data of the dislocation, the velocity of the dislocation segment is calculated as $\Delta x/\Delta t$ where the distance $\Delta x$ that a dislocation segment has travelled during $\Delta t$ is perpendicular on the initial line (shown as the solid black line in Fig.~\ref{fig2}b. In our calculations, we guarantee the time span between $t_{0}$ and $t_{1}$ being small enough to be able to calculate $\Delta x$. Only at the end points of the polygon $\Delta x$ is calculated directly as the distance between the dislocation end point, as shown by the green dashed line in Fig.~\ref{fig2}b. A calculation based on two subsequent frames is shown in Fig.\,\ref{fig5}c.} There, we can see that the velocity also varies along the dislocation line. Interestingly, we see that around $P_{3}$, it clearly shows that the velocity on the left side (already passed) of $P_{3}$ is larger than on the right side (prior to passing) of $P_{3}$. This observation is consistent with the conventional picture of dislocation depinning by quenched disorder: in the pinning regime, the velocity of the dislocation is slowed down while in the depinning regime, the sudden increase in velocity leads to the so-called dislocation avalanche.

	\begin{figure}[ht!]%
		\centering
		\includegraphics[width=\textwidth]{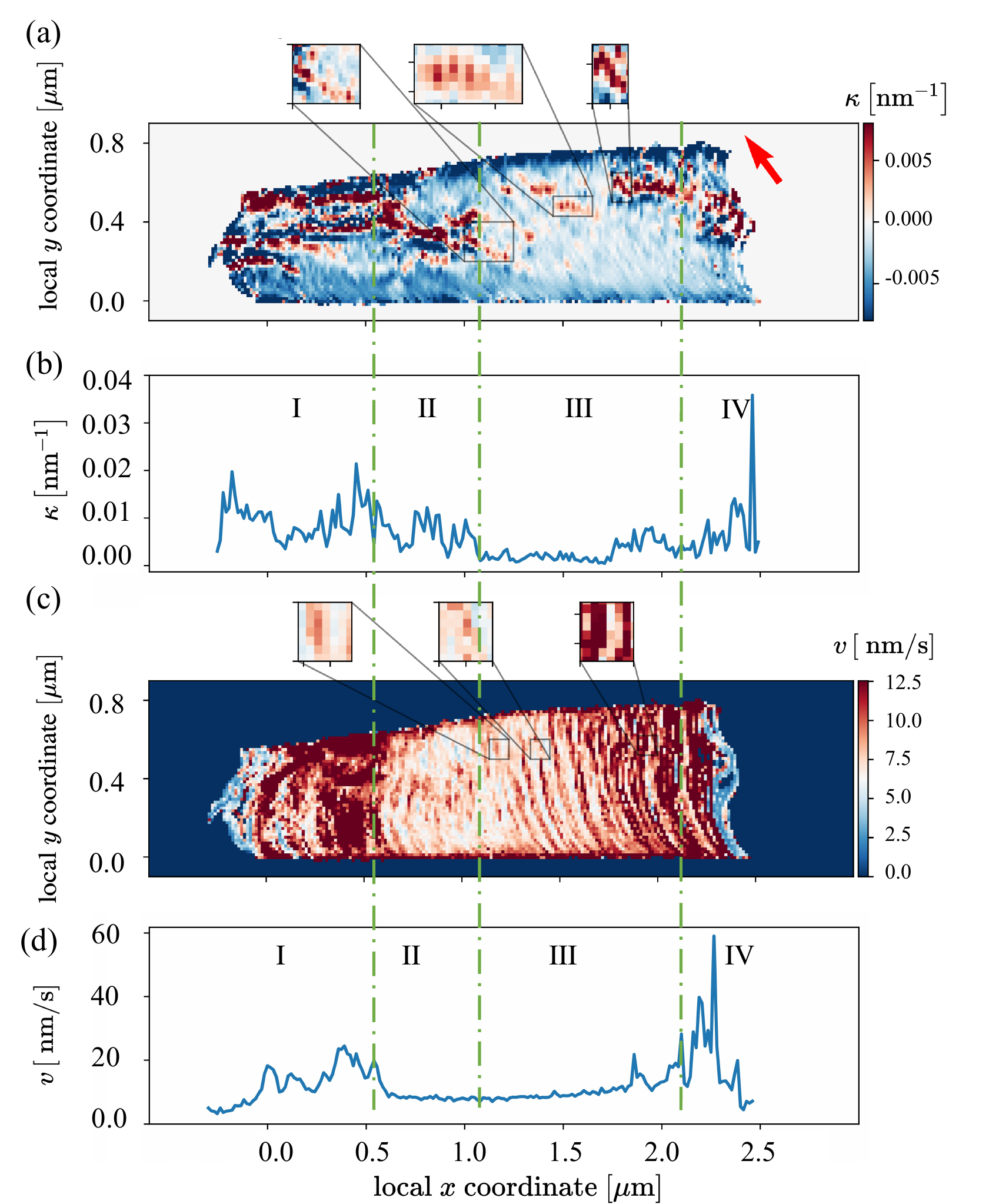}
		\caption{Spatio-temporal averaging of curvature and velocity based on 303 images. (a) distribution of curvature with a pixel size of $15\,\mathrm{nm}$. The red arrow represents the direction of the Burgers vector $\left(-1,1, 0\right)$. (b) The distribution of the averaged (along $y$ direction) curvature along the pile-up direction ($x$-axis). (c) Distribution of velocity with a pixel size of $15\,\mathrm{nm}$. (d) The distribution of the averaged (along $y$ direction) velocity along the pile-up direction ($x$-axis).}\label{fig6}
	\end{figure}	

	Fig.~\ref{fig6}a shows the `scanning' results of the curvature based on all 303 frame images, while Fig.~\ref{fig6}c shows the `scanning' results of the velocity. It can be seen in Fig.~\ref{fig6}a that at the head of the pile up (left part of the slip plane), there are several bands of high curvature regions. These bands match the low-velocity bands in Fig.~\ref{fig6}c indicating the existence of local pinning points. Away from this pinning point-rich area, there are also some high curvature bands highlighted by the magnified regions in other locations on the slip plane. Interestingly, the direction of these bands is close to the direction of the Burgers vector.  Among the three highlighted bands, the one on the right is the location where $P_{3}$ is defined and studied in detail in the previous section. The formation of a band around $P_{3}$ may have two possible reasons: 1) there are multiple local pinning points lining up. However, in the TEM video, we did not see the consecutive pinning when a single dislocation travels in that region. 2) the local pinning point is sheared and dragged by the dislocation, which, therefore, moves towards the direction of the Burgers vector. The second reason seems highly possible in the sense that the physical origin of the local pinning point in HEA is the local ordering. When a dislocation passes by, the atoms above and below the slip plane are relatively shifted by the magnitude of one Burgers vector, therefore altering the local ordering along the direction of the Burgers vector. This phenomenon is different from the conventional solute hardening where the solute (pinning point) is believed to be immobile, yet different from the Cottrell drag where solutes are mobile enough to keep up with moving dislocations. This observation also excludes the possibility of dislocation getting pinned by jogs whose position won't change.
	
	The coarse graining carried out with small voxels reflects the local features of the dislocation. In Fig.\,\ref{fig6}b and Fig.\,\ref{fig6}d, we averaged the scanned curvature and velocity along the $y$ direction, thus observing how the average curvature and velocity vary along the $x$ direction. Our probe for sampling the slip plane is a dislocation pile-up structure (even though the particular structure is not ideal since the leading dislocation is not entirely blocked).  For an ideal dislocation pile-up, it is expected that the velocity of dislocations decreases from the tail to the head of the pileup. Any deviation (from this expectation) of our observation would result from the local pinning effect.
In Fig.\,\ref{fig6}b, it can be seen that the there are many high curvature regions along this pile-up, indicating the common existence of the local pinning effect. The local pinning effect results in a very different velocity profile compared to an ideal pile-up, as shown in Fig.~\ref{fig6}d. An interesting observation is highlighted as region II, where the velocity is almost a constant value even though the curvature value is large and fluctuates. In general, close to an isolated pinning point, it is expected that there should be a fluctuation in the velocity because of the pinning-depinning mechanism (one can think of a rubber band release from the pinning state), as shown on the right of region III and the left of region IV, can also be seen in Fig.~\ref{fig6}c. The low velocity variations in region II indicates that this is probably a pinning point-rich region where dislocation is frequently pinned (they get pinned almost immediately after depinned from the previous pinning site). In principle, our method can also identify each single pining point in such region, however, this would require a much better resolution both spatially and temporally (limited by the experimental video frame rate).

\section{Conclusion}\label{sec4}
	In this paper, we investigated the pinning point dynamics of the equimolar CoCrFeMnNi HEA at near liquid nitrogen temperature using in situ TEM straining. For this, we introduced a new data mining approach that relies on spatio-temporal coarse graining of pile-up dislocations used as probes of the local energy landscape. Such analysis of the morphology and motion of dislocations leads to the following conclusions:
	
	\begin{enumerate}
		\item {The strength of the pinning obstacles changes upon dislocations passing by. The change in strength is random as some obstacles get weakened while others get strengthened.}		
		\item {The spatial location of the pinning obstacles is observed to vary along the Burgers vector direction. }		
		\item {The spatial distribution of the pinning obstacles is not homogeneous: there are isolated pinning obstacles and close-packed pinning obstacles}.
		
		\item {In the specific case of HEAs, this approach establishes for the first time that moving dislocations experience a friction arising from obstacles that have random strength and distribution and that this randomness is shifted in time and space by passing dislocations; the friction effect is not completely annihilated.}
	\end{enumerate}
	Our data mining approach -- using dislocations as probe, and the coarse-grained field quantities as resulting `scanning'  signals -- is able to detect the pinning point locations and strength. It is thus  a very promising technique for further fundamental in-situ TEM studies of dislocation microstructures and the related property of materials.

	\section*{Declaration of Competing Interest}
	The authors declare that they have no known competing financial interests or personal relationships that could have appeared to influence the work reported in this paper.

	\section*{Acknowledgments}
	This work was funded by the European Research Council through the ERC Grant Agreement No. 759419 MuDiLingo (``A Multiscale Dislocation Language for Data-Driven Materials'') and supported partly by CEMES-CNRS and partly by the European Union's horizon 2020 research and innovation program under grant agreement n$^\circ$823717 (ESTEEM3) for the microscopy facilities.

\bibliography{els-article}
	
\end{document}